\documentclass[aip,apl,reprint]{revtex4-2}

\usepackage{graphicx}   
\usepackage{dcolumn}    
\usepackage{bm}         
\usepackage{siunitx}    
\DeclareSIUnit{\mbar}{mbar}
\usepackage[colorlinks=true, linkcolor=blue, citecolor=blue, urlcolor=blue]{hyperref} 
\usepackage{xcolor}

\draft 

\begin{document}


\title{Surface-Localized Magnetic Order in RuO$_2$ Thin Films Revealed by Low-Energy Muon Probes} 



\author{Akashdeep Akashdeep}
\affiliation{Institute of Physics, Johannes Gutenberg-University Mainz, 55128 Mainz, Germany}

\author{Sachin Krishnia}
\affiliation{Institute of Physics, Johannes Gutenberg-University Mainz, 55128 Mainz, Germany}

\author{Jae-Hyun Ha}
\affiliation{Center for Spintronics, Korea Institute of Science and Technology (KIST), Seoul, Republic of Korea}

\author{Siyeon An}
\affiliation{Center for Spintronics, Korea Institute of Science and Technology (KIST), Seoul, Republic of Korea}

\author{Maik Gaerner}
\affiliation{Faculty of Physics, Bielefeld University, 33615 Bielefeld, Germany}

\author{Thomas Prokscha}
\affiliation{PSI Center for Neutron and Muon Sciences, 5232 Villigen PSI, Switzerland}

\author{Andreas Suter}
\affiliation{PSI Center for Neutron and Muon Sciences, 5232 Villigen PSI, Switzerland}

\author{Gianluca Janka}
\affiliation{PSI Center for Neutron and Muon Sciences, 5232 Villigen PSI, Switzerland}

\author{Günter Reiss}
\affiliation{Faculty of Physics, Bielefeld University, 33615 Bielefeld, Germany}

\author{Timo Kuschel}
\affiliation{Institute of Physics, Johannes Gutenberg-University Mainz, 55128 Mainz, Germany}
\affiliation{Faculty of Physics, Bielefeld University, 33615 Bielefeld, Germany}

\author{Dong-Soo Han}
\affiliation{Center for Spintronics, Korea Institute of Science and Technology (KIST), Seoul, Republic of Korea}

\author{Angelo Di Bernardo}
\affiliation{Department of Physics, University of Konstanz, Universitaetsstrasse 10, 78464 Konstanz, Germany}
\affiliation{Department of Physics, University of Salerno, Via Giovanni Paolo II 132, 84084 Fisciano SA, Italy}

\author{Zaher Salman}
\affiliation{PSI Center for Neutron and Muon Sciences, 5232 Villigen PSI, Switzerland}

\author{Gerhard Jakob}
\affiliation{Institute of Physics, Johannes Gutenberg-University Mainz, 55128 Mainz, Germany}

\author{Mathias Kläui}
\email[]{klaeui@uni-mainz.de}
\affiliation{Institute of Physics, Johannes Gutenberg-University Mainz, 55128 Mainz, Germany}
\affiliation{Department of Physics, Center for Quantum Spintronics, Norwegian University of Science and Technology, 7491 Trondheim, Norway}


\date{\today}

\begin{abstract}

Ruthenium dioxide (RuO$_2$) has recently emerged as a altermagnetic candidate, but its intrinsic magnetic ground state in thin films remains widely debated. This study aims to clarify the nature and spatial extent of the magnetic order in RuO$_2$ thin films grown under different conditions. Thin films of RuO$_2$ with thicknesses of \SIlist{30;33}{\nano\meter} are deposited by pulsed laser deposition and sputtering onto TiO$_2$(110) and Al$_2$O$_3$(\={1}102) substrates, respectively. Low-energy muon spin rotation/relaxation (LE-$\mu$SR) with depth-resolved sensitivity measurements are performed in transverse magnetic fields (TF) from \SIrange{4}{290}{\kelvin}. The $\mu$SR data collected with muon implantation energy of \SI{1}{\kilo\eV} reveal that magnetic signals originate from the near-surface region of the film ($\lesssim$\SI{10}{\nano\meter}), and the affected volume fraction is of approximately 8.5\%. The localized magnetic response is consistent across different substrates, growth techniques, and parameter sets, suggesting a common origin related to surface defects and dimensionality effects. The combined use of TF-$\mu$SR and study of depth-dependent implantation with low-energy muons provides direct evidence for surface-confined, inhomogeneous static magnetic order in RuO$_2$ thin films, helping reconcile discrepancies. These findings underscore the importance of considering reduced-dimensional contributions and motivate further investigation into the role of defects, strain, and stoichiometry on the magnetic properties of RuO$_2$, especially at the surface.

\end{abstract}

\pacs{}

\maketitle 

Magnetic materials are conventionally categorized into ferromagnets (FMs) and antiferromagnets (AFMs), distinguished by the presence or absence of momentum-independent spin splitting in collinear spin configurations. Recently, a class of magnetic materials, termed \emph{altermagnets}, has been identified, exhibiting properties that transcend this traditional classification. Altermagnets display momentum-dependent spin splitting, possess compensated magnetic order similar to AFMs, and exhibit spin-split electronic bands characteristic of FMs \cite{Smejkal2022_NatRevMat, Smejkal2022_PhysRevX, PhysRevLett.126.127701}. The spin and spatial alternation of the magnetic sublattices in altermagnets arises from the intrinsic crystal and magnetic symmetries, which are invariant under a combination of spatial rotation and spin rotation operations. This unique symmetry allows momentum-dependent spin-splitting in the electronic bands, even in the absence of net magnetization. These features enable the generation of spin currents in altermagnets without requiring spin--orbit coupling or uncompensated magnetic moments, offering new opportunities for spintronic applications \cite{Smejkal2022_NatRevMat, Smejkal2022_PhysRevX, PhysRevLett.126.127701}.

\begin{figure}[h]
    \centering
    \includegraphics[width= \columnwidth]{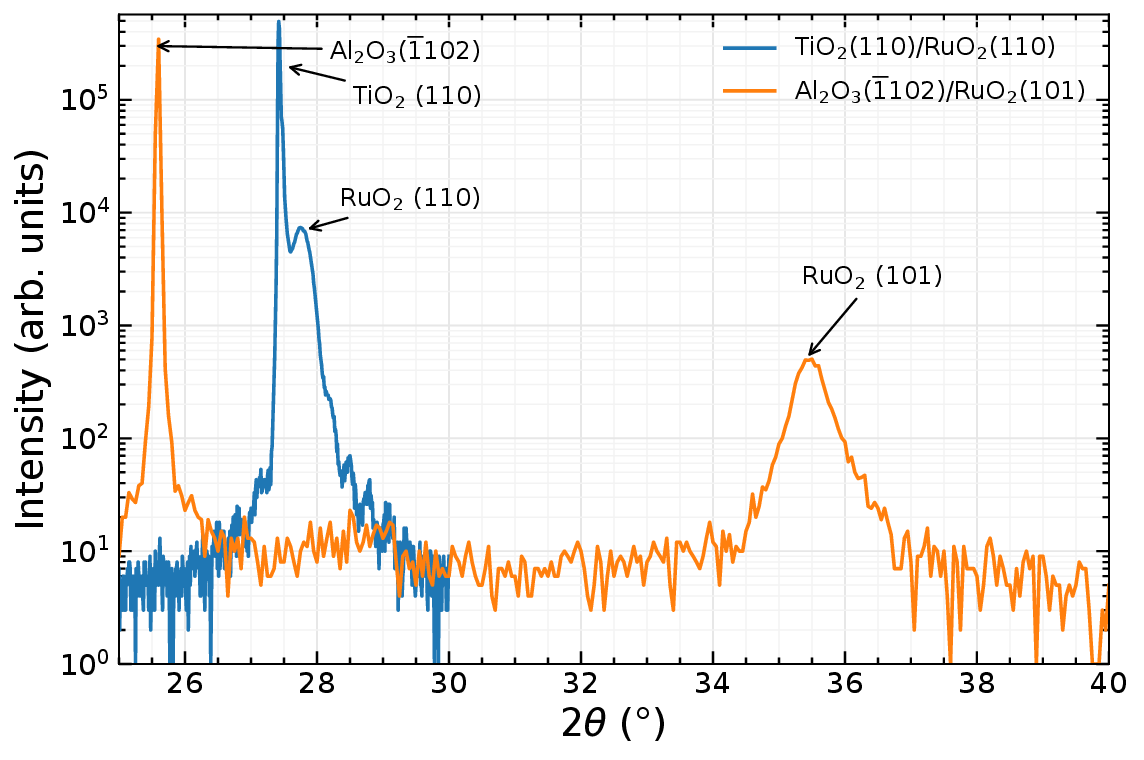}
    \caption{XRD patterns measured with the scattering vector oriented normal to the (110)-oriented rutile TiO$_2$ in blue and the (\={1}102)-oriented Al$_2$O$_3$ substrate in orange. }
    \label{fig:XRD_Comparison}
\end{figure}

Ruthenium dioxide (RuO$_2$) has attracted particular attention, as it crystallizes in the rutile structure and has been reported to host a compensated magnetic order with an antiparallel sublattice \cite{PhysRevLett.118.077201, PhysRevLett.122.017202, PhysRevB.99.184432}. In addition, strong spin splitting in its electronic band structure has been directly observed by angle-resolved photoemission spectroscopy (ARPES) \cite{doi:10.1126/sciadv.adj4883, lin2024observationgiantspinsplitting}, along with diverse altermagnetic features via different means, including the anomalous Hall effect \cite{Feng2022_NatElectron, 10.1063/5.0160335, wang2023emergent}, charge-to-spin conversion \cite{PhysRevLett.128.197202, PhysRevLett.129.137201, Bose2022, https://doi.org/10.1002/advs.202400967}, spin-to-charge conversion \cite{PhysRevLett.130.216701, https://doi.org/10.1002/advs.202400967, PhysRevLett.133.046701, z478-fgl8}, magneto-optical effects \cite{He2025, jeong2025altermagneticpolarmetallicphase, weber2024opticalexcitationspinpolarization}, terahertz emission \cite{https://doi.org/10.1002/adom.202300177, 10.1117/1.AP.5.5.056006, z478-fgl8}, and tunneling magnetoresistance junctions \cite{nrk5-5zrj}.

Despite these findings, the intrinsic magnetic ground state of RuO$_2$ remains under debate. Several recent experimental and theoretical studies suggest that RuO$_2$ is non-magnetic in its ground state \cite{PhysRevLett.132.166702, PhysRevB.111.134450, Kiefer_2025, PhysRevB.110.064432, PhysRevB.111.L041115, yumnam2025constraintsmagnetismcorrelationsruo2, qian2025determiningnaturemagnetismaltermagnetic, osumi2025spindegeneratebulkbandstopological, peng2025universal, 5js8-2hj8, kessler2024absence, keßler2025moireassistedchargeinstabilityultrathin,PhysRevApplied.23.054082, PhysRevLett.133.176401, abel2025probingmagneticpropertiesruo2, jechumtál2025spintochargecurrentconversionaltermagneticcandidate, PhysRevB.109.134424}. Most of these studies have focused on bulk single crystals, with only a few also investigating thin films \cite{kessler2024absence, PhysRevApplied.23.054082, PhysRevLett.133.176401, abel2025probingmagneticpropertiesruo2, jechumtál2025spintochargecurrentconversionaltermagneticcandidate, keßler2025moireassistedchargeinstabilityultrathin}. For bulk crystals, the prevailing view is that the ground state is non-magnetic. In contrast, the situation for thin films remains unresolved, as experimental reports provide conflicting results. It is, however, widely recognized that lattice strain, defects, interface, and other growth-related factors in thin films can induce unusual magnetic and transport properties \cite{PhysRevB.109.134424, ko2018understanding, PhysRevB.106.195135, https://doi.org/10.1002/celc.202300659, PhysRevLett.133.176401, abel2025probingmagneticpropertiesruo2, jechumtál2025spintochargecurrentconversionaltermagneticcandidate, 4m5d-ylyr, PhysRevLett.125.147001, ruf2021strain, doi:10.1073/pnas.2500831122, 6fxv-153y, Brahimi_2025}.

The conflicting results in thin films underscore the need for a detailed investigation into the magnetic properties of RuO$_2$ thin films. Earlier work using muon spin spectroscopy on \SI{10}{\nano\meter} thin films was limited by the low thickness, and the absence of a significant magnetic signal made interpretation of the results challenging \cite{kessler2024absence}. In the present study, we investigate thin films of \SIlist{30;33}{\nano\meter} grown under different conditions on different substrates using low-energy muon spin spectroscopy, enabling depth-resolved measurements of magnetic moment with high sensitivity \cite{BATOR2012137}.


Thin films of RuO$_2$ were prepared to investigate their magnetic properties in the context of altermagnetism. Four RuO$_2$(110) films, each \SI{33}{\nano\meter} thick, were grown on TiO$_2$(110) substrates by pulsed laser deposition (PLD) under conditions: oxygen pressures of \SI{0.025}{\mbar}(\(\pm 0.005\)), and substrate temperatures of \SI{375}{\celsius}(\(\pm 25\)). Additionally, four RuO$_2$(101) films, each \SI{30}{\nano\meter} thick, were deposited on Al$_2$O$_3$(\={1}102) substrates by reactive DC magnetron sputtering under a pressure of \SI{0.004}{\mbar}, with a mixture of $\mathrm{O}_2/(\mathrm{Ar}+\mathrm{O}_2)=0.167$ and substrate temperatures of \SI{700}{\celsius}. The crystalline quality is verified by X-ray diffraction (XRD), and reflection high-energy electron diffraction (RHEED) was additionally employed for the PLD-grown films. Each sample has a surface area of about \SI{1}{\centi\meter}$\times$\SI{1}{\centi\meter}. Both sets of four samples on the same substrate were mounted on a nickel-coated aluminum plate to maximize the incident muon beam cross-section and improve counting statistics, resulting in a total area of \SI{2}{\centi\meter}$\times$\SI{2}{\centi\meter}.


Low-energy muon spin rotation/relaxation (LE-$\mu$SR) experiments were conducted on the $\mu$E4 beamline at the Swiss Muon Source (S$\mu$S), Paul Scherrer Institute (PSI), Switzerland, using the low-energy muon (LEM) spectrometer \cite{Prokscha2008NIMA}. Positive muons ($\mu^+$), produced via pion decay from proton–carbon collisions, were nearly fully spin-polarized at the source \cite{NI2023168399, hillier2022muon, BATOR2012137}. The muons were slowed down via moderators to energies of tens of eV, and then electrostatically accelerated to implantation energies between \SI{1}{\kilo\eV} and \SI{12}{\kilo\eV}, enabling depth-resolved magnetic measurements \cite{Morenzoni1994PRL}. Measurements were performed in a transverse magnetic field (TF) of either \SI{10}{\milli\tesla} or \SI{140}{\milli\tesla}, applied perpendicular to both the initial muon spin polarization and the sample surface. The temperature was varied from \SI{4}{\kelvin} to \SI{290}{\kelvin}. For the RuO$_2$(110)/TiO$_2$(110) films, the muon spin was oriented in-plane and normal to the $c$-axis. For the RuO$_2$(101)/Al$_2$O$_3$(\={1}102) films, the muon spin was oriented in-plane and 55.3° to the $c$-axis. Monte Carlo TRIM.SP simulations were used to estimate the mean muon implantation depth and distribution in the samples \cite{TRIMSP, musruserwebsite}, as shown in Supplementary Figures S1 and S2. Muon spin rotation and relaxation were monitored by detecting decay positrons, preferentially emitted along the muon spin direction. Symmetrically arranged detectors around the sample recorded the events, with approximately $3\times 10^6$ counts per data point to ensure high statistical accuracy.

\begin{figure*}[t]
    \centering
    \includegraphics[width=\textwidth]{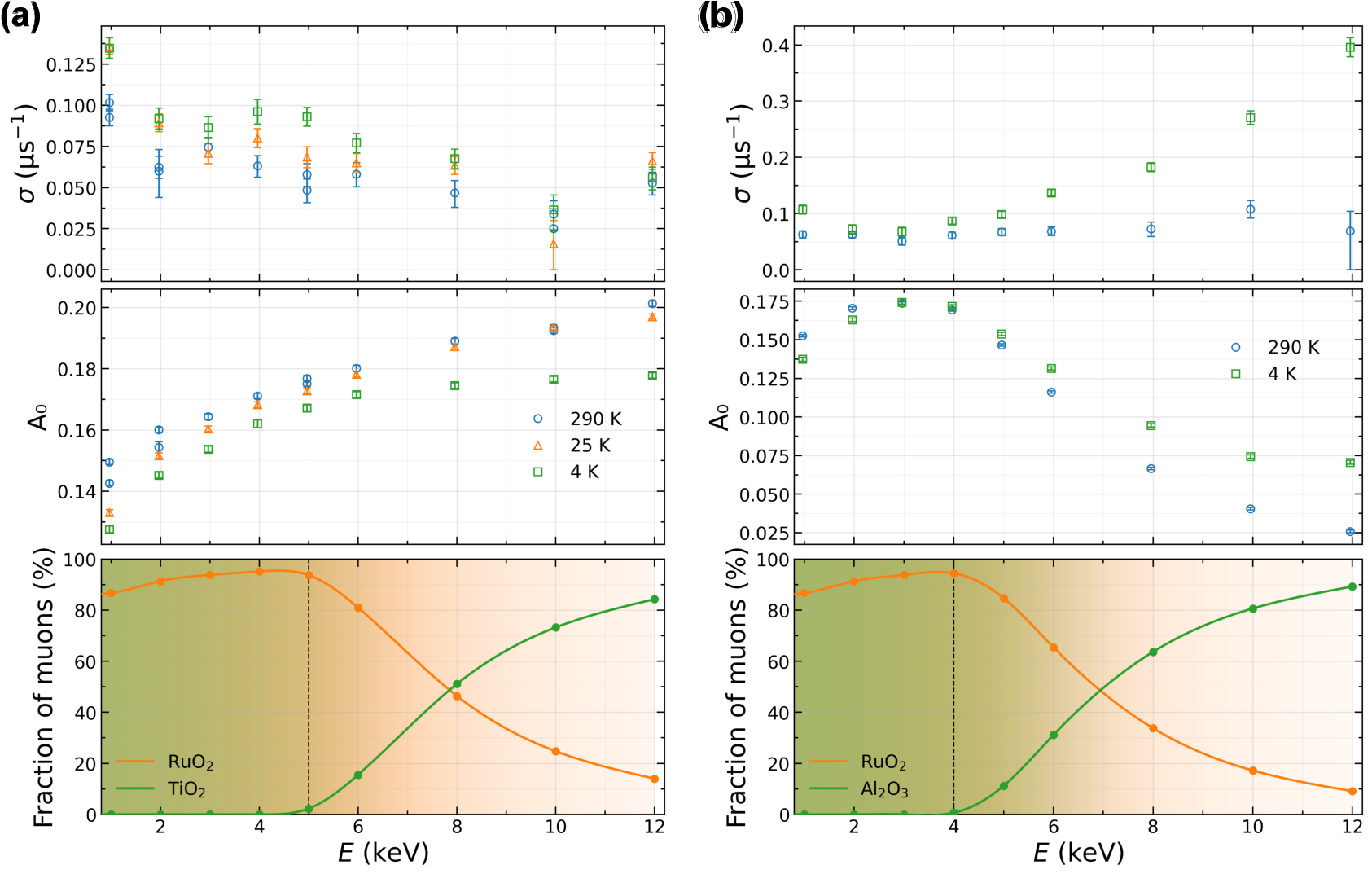}
    \caption{Dependence of muon spin damping rate $\sigma$, initial asymmetry $A_0$,  and fraction of muons stopping in the thin film and substrate on implantation energy at different temperatures for (a) RuO$_2$(110)/TiO$_2$(110) films (b) RuO$_2$(101)/Al$_2$O$_3$(\={1}102) films at applied magnetic field of \SI{10}{\milli\tesla}. (c) The green and orange shaded areas represent the fraction of muons in the RuO$_2$ and substrate layers, respectively.}
    \label{fig:Figure 2_ParametersVsE}
\end{figure*}


The $\mu$SR time spectra were analyzed using the \textsc{Musrfit} software package \cite{suter2012musrfit, musruserwebsite}. The asymmetry signal was fitted with a damped cosine function with a Gaussian envelope, 
$A(t) = A_0 \cos(\gamma_\mu B t + \varphi)\, e^{-\frac{1}{2} (\sigma t)^2}$, to extract the muon spin damping rate $\sigma$, the local magnetic field $B$, and the initial asymmetry $A_0$. Here, $\gamma_\mu$ is the muon gyromagnetic ratio and $\varphi$ is the initial phase offset.


\begin{figure*}[t]
    \centering
    \includegraphics[width=\textwidth]{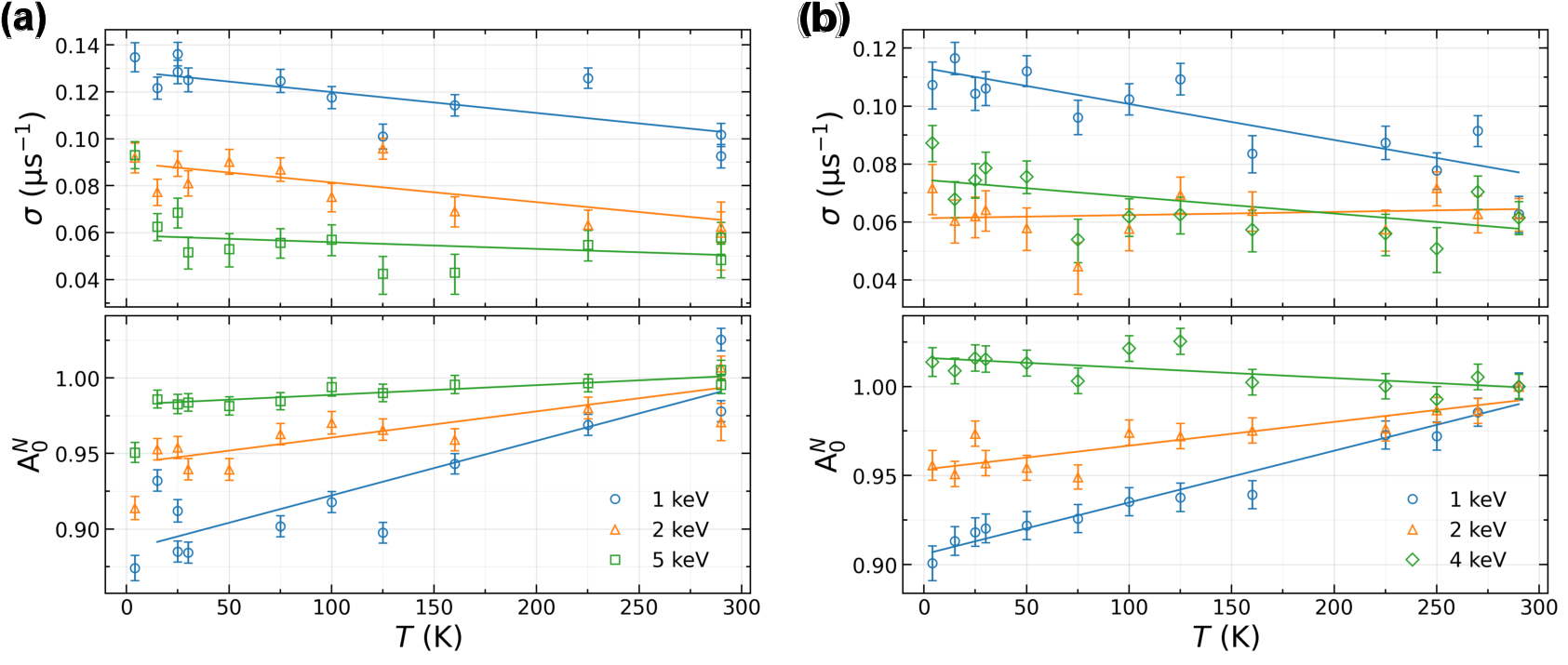}
    \caption{Temperature dependence of muon spin damping rate $\sigma$ and normalized initial asymmetry $A_0^{N}$ films at different implantation energies for (a) RuO$_2$(110)/TiO$_2$(110) films and (b) RuO$_2$(101)/Al$_2$O$_3$(\={1}102) films at applied magnetic field of \SI{10}{\milli\tesla}. }
    \label{fig:Figure 3_ParametersVsT}
\end{figure*}



Figure~\ref{fig:XRD_Comparison} shows the X-ray diffraction (XRD) patterns measured with the scattering vector normal to the (110)-oriented rutile TiO$_2$ and the (\={1}102)-oriented Al$_2$O$_3$ substrate. Both RuO$_2$(110)/TiO$_2$(110) and RuO$_2$(101)/Al$_2$O$_3$(\={1}102) thin film samples show an XRD pattern corresponding to a single crystallographic orientation of the deposited layer. TiO$_2$(110)/RuO$_2$(110) even exhibits Laue oscillations, a signature of a single-crystalline thin film with sharp interfaces \cite{MillerLemonChoffelRichHarvelJohnson+2022+313+322}. 
Additionally, the RHEED pattern for the TiO$_2$(110)/RuO$_2$(110) sample is presented in the supplemental material (Supplementary Figure S3). The modulated streaks in the RHEED pattern corroborate the single-crystalline growth and a surface roughness of only a few atomic layers.


TF LE-$\mu$SR measurements were conducted with implantation energies ranging from \SI{1}{\kilo\eV} to \SI{12}{\kilo\eV} to probe depth-dependent magnetic behavior in both RuO$_2$ thin-film systems. The extracted parameters included $\sigma$ and $A_0$. Representative results between \SI{4}{\kelvin} and \SI{290}{\kelvin} are shown for RuO$_2$ films grown on TiO$_2$ and Al$_2$O$_3$ substrates in Figures~\ref{fig:Figure 2_ParametersVsE}(a) and \ref{fig:Figure 2_ParametersVsE}(b), respectively.

For the RuO$_2$/TiO$_2$ films, low implantation energies (e.g., \SI{1}{\kilo\eV}, corresponding to muons stopping primarily in the top $\sim$\SI{10}{\nano\meter} of the film) resulted in a pronounced and gradual increase in $\sigma$ and a significant gradual decrease in $A_0$ at \SI{25}{\kelvin} compared to \SI{290}{\kelvin}. In contrast, higher implantation energies (e.g., \SI{12}{\kilo\eV}, corresponding to muons stopping predominantly in the substrate) showed much weaker temperature dependence. An additional anomaly was observed as a more substantial reduction in $A_0$ at \SI{4}{\kelvin} for high implantation energies. This behavior is attributed to the paramagnetic muonium complex signal, which forms below 10~K in rutile TiO$_2$ \cite{PhysRevB.92.075203, PhysRevB.92.081202}.

For the RuO$_2$/Al$_2$O$_3$ films, low implantation energies (\SIrange{1}{2}{\kilo\eV}) lead to an increase in $\sigma$ and a decrease in $A_0$ at \SI{4}{\kelvin} relative to \SI{290}{\kelvin}. These changes are less pronounced at intermediate implantation energies (\SIrange{3}{4}{\kilo\eV}). At higher implantation energies, i.e., for which the muons stop primarily in the Al$_2$O$_3$ substrate, a more substantial increase in $\sigma$ and a stronger decrease in $A_0$ were observed. In Al$_2$O$_3$, this signal is due to the muon experiencing the dipolar field of Al nuclear moments, while most of the lost asymmetry is due to muonium formation in this material \cite{PhysRevB.99.064423, PhysRevB.96.184402, PhysRevB.103.125202, PhysRevLett.98.227401}.

Overall, the data for RuO$_2$ grown on both substrates demonstrate that the main changes in $\sigma$ and $A_0$ occur at low implantation energies, confirming that the magnetic response originates from the near-surface regions of the RuO$_2$ thin films. Additional measurements were conducted in a TF of \SI{140}{\milli\tesla} on both samples, yielding trends that are qualitatively analogous to the \SI{10}{\milli\tesla} measurements. These additional measurements are presented in the supplemental material provided. The variations in $B$ are within the systematic error limits of the measurement apparatus, indicating that there is no observable temperature or implantation energy dependence in $B$.


Detailed temperature-dependent TF-$\mu$SR measurements were performed between \SI{4}{\kelvin} and \SI{290}{\kelvin} to explore the magnetic response further. Figures~\ref{fig:Figure 3_ParametersVsT}(a) and \ref{fig:Figure 3_ParametersVsT}(b) illustrate the evolution of $\sigma$ and $A_0^{N}$
(normalized $A_0$ with respect to weighted mean of $A_0$(290 K) values) with different muon implantation energies for RuO$_2$ films grown on TiO$_2$ and Al$_2$O$_3$ substrates, respectively. We observe a gradual increase in $\sigma$ and a corresponding decrease in $A_0^{N}$ with decreasing temperature, as indicated by the linear fits. For the RuO$_2$(110)/TiO$_2$(110) films, the \SI{4}{\kelvin} data points were excluded from the fits due to the dominant paramagnetic contribution from the TiO$_2$ substrate \cite{PhysRevB.92.075203}.

As the temperature decreases, $\sigma$ increases while $A_0^{N}$ decreases for both samples, with the most substantial variation observed at \SI{1}{\kilo\eV} implantation energy. At higher implantation energy, for which most of the muons stop deeper in the thin film and partially in the substrates, we detect a weak variation of $\sigma$ and $A_0^{N}$  with temperature in both samples. The observed temperature dependence in RuO$_2$ strongly suggests that the magnetic response originates in the near-surface region (approximately \SI{10}{\nano\meter}) of the RuO$_2$ thin films. The relative change of the initial asymmetry \( A_0 \) between 290 K and 4 K measured at 1~keV implantation energy, as \((A_0(290~\mathrm{K}) - A_0(4~\mathrm{K}))/A_0(290~\mathrm{K})\), indicates that $10.5\% \pm 3.6\%$ for TiO$_2$/RuO$_2$ and $8.4\% \pm 0.9\%$ for Al$_2$O$_3$/RuO$_2$ (95\% confidence interval) of the near-surface volume exhibits inhomogeneous magnetic order. The gradual temperature dependence of $\sigma$, $A_0^{N}$, and the small near-surface volume with a magnetic feature suggest that the underlying magnetic response is characterized by inhomogeneous static magnetic order on the surface of the thin film, with no evidence for long-range magnetic order inside the thin film. Additional results obtained in a TF of \SI{140}{\milli\tesla} are provided in the supplemental material (Supplementary Figures S4 and S5).

Temperature-dependent TF-$\mu$SR measurements revealed that RuO$_2$ thin films exhibited inhomogeneous static magnetic order confined to the near-surface region. Depth-resolved measurements, enabled by varying muon implantation energies, showed systematic changes in the $\sigma$ and $A_0$ near the surface, which became more pronounced at lower temperatures. This behavior suggests that surface-related effects, rather than bulk phenomena, are the primary drivers of the observed magnetic response. Quantitative analysis of the TF data indicated that the magnetic volume fraction is comparatively small, with a weighted mean of approximately $\sim$8.5\% within the top $\sim$\SI{10}{\nano\meter} of the films. This surface-localized magnetism may help to explain the manifestation of possible altermagnetic-like or other anomalous features in thin films, in contrast to bulk single crystals of RuO$_2$, which are widely reported to be non-magnetic \cite{kessler2024absence, PhysRevLett.132.166702, PhysRevB.109.134424}. Indeed, our other thin films have been confirmed as prototypical altermagnetic materials using surface-sensitive techniques, and microscopic evidence supports this classification. This evidence includes the direct observation of strong time-reversal symmetry breaking in the electronic band structure, as revealed by magnetic circular dichroism (MCD) in ARPES \cite{doi:10.1126/sciadv.adj4883}. The d-wave symmetry of the altermagnetic phase in ultrathin films has also been confirmed through a unique response to linearly polarized light. This response shows excitation-angle-dependent transient spin polarization via time-resolved magneto-optic Kerr effect (TR-MOKE) \cite{weber2024opticalexcitationspinpolarization}.

The comparison between thin films grown via PLD and sputtering demonstrates that the surface magnetism is robust against variations in substrates, growth technique, and parameters, suggesting a common underlying mechanism. Our results are consistent with the prior $\mu$SR study of RuO$_2$ thin films \cite{kessler2024absence}, which reported that films have stronger damping rates than the bulk. Notably, due to limited depth resolution, earlier work on \SI{10}{\nano\meter} films could not establish a direct link to surface magnetism \cite{kessler2024absence}. By employing a \SI{30}{\nano\meter}-thick film for which muon implantation beyond the surface layer is possible, our results directly confirm that the static weak magnetic fields are confined to the $\sim$8.5\% of top $\sim$\SI{10}{\nano\meter}. These findings highlight the potential role of dimensionality and surface defects in modifying the spin dynamics of RuO$_2$ thin films. The confinement of magnetism to the surface raises essential questions about whether the observed behavior is intrinsic altermagnetism or arises from strain, interface defect-induced, and localized states. However, the interpretation of intrinsic surface magnetism is challenged by high-resolution studies; specifically, spin-polarized scanning tunneling microscopy (SP-STM) measurements on atomically ordered ultrathin (110) surfaces found no evidence of magnetic contrast \cite{keßler2025moireassistedchargeinstabilityultrathin}. This suggests that while RuO$_2$ thin films exhibit inhomogeneous static magnetic order detected by $\mu$SR, it may be below the SP-STM detection limit or originate from a defect rather than intrinsic surface electronic order. The results underscore the importance of considering surface contributions when interpreting magnetic measurements in reduced-dimensional systems.

While TF-$\mu$SR combined with depth-dependent implantation provided valuable insights, the technique inherently averages over different crystallographic orientations and momentum directions, limiting the ability to unambiguously distinguish between intrinsic altermagnetic order and defect-related anomalies \cite{PhysRevB.109.134424, ko2018understanding, PhysRevB.106.195135, https://doi.org/10.1002/celc.202300659, PhysRevLett.133.176401}. Resolving this distinction requires complementary approaches that offer different levels of resolution: while methods like MCD-ARPES confirm the intrinsic altermagnetic electronic structure and TR-MOKE confirm dynamic altermagnetic behavior, surface-sensitive probes such as the SP-STM are necessary, which could help to resolve the microscopic origin of the surface magnetism.

In this study, we employed depth-resolved LE-$\mu$SR to investigate the magnetic properties of RuO$_2$ thin films grown by PLD and sputtering. We probed varying depth regions within the films by changing the muon implantation energy. We demonstrated that the observed magnetic response is confined to the near-surface layer, with no evidence of long-range magnetic order inside the thin film. These findings help to  reconcile discrepancies in the literature regarding the magnetism of RuO$_2$ thin films, offering evidence that surface-related defects and dimensionality effects play a significant role in the emergence of weak magnetic signals. While the combined use of TF-$\mu$SR and depth-dependent implantation provides high sensitivity to local magnetic environments, further investigations using complementary surface-sensitive techniques are needed to identify the microscopic origin of the observed magnetism and to determine whether it is intrinsic altermagnetism or surface defect-induced. Overall, this work establishes a clear experimental basis for the role of surface effects in RuO$_2$ thin films and highlights promising avenues for future exploration in spintronic materials with reduced dimensionality.


%
%

%


\section*{Supplementary Material}
See supplementary material for additional TRIM.SP simulations, RHEED patterns, and muon spectroscopy data at \SI{140}{\milli\tesla}.

\section*{Data Availability}
The data supporting the findings of this study are openly available in Zenodo \cite{akashdeep_2025_17868066}. 

\begin{acknowledgments}
All authors from Mainz also gratefully acknowledge funding support from the Deutsche Forschungsgemeinschaft (DFG) under the framework of the Collaborative Research Center TRR 173–268565370 Spin+X (Project B02, A01, and A12), TRR 288-422213477 Elasto-Q-Mat (Project A12), and National Research Council of Science and Technology (NST) grant by the Korean government MSIT (Grant No.GTL24041-000. All authors from KIST also express their profound gratitude to the ASTRA Project, which is funded by the National Research Foundation (NRF) and financially supported by the Ministry of Science and ICT (RS-2024-00488149). The LE-$\mu$SR experiments were performed at the Swiss Muon Source S$\mu$S, Paul Scherrer Institute, Villigen, Switzerland. The authors declare no conflicts of interest.
\end{acknowledgments}


\bibliographystyle{aipnum4-2} 
\bibliography{main_bib} 

\end{document}